# Neuromorphic Computing through Time-Multiplexing with a Spin-Torque Nano-Oscillator

M. Riou[1], F. Abreu Araujo[1], J. Torrejon[1], S. Tsunegi[2], G. Khalsa[3], D. Querlioz[4], P. Bortolotti[1], V. Cros[1], K. Yakushiji[2], A. Fukushima[2], H. Kubota[2], S. Yuasa[2], M. D. Stiles[3], and J. Grollier[1]
[1]Unité Mixte de Physique, CNRS, Thales, Univ. Paris-Sud, Université Paris-Saclay, France, email: julie.grollier@cnrs-thales.fr
[2] National Institute of Advanced Industrial Science and Technology (AIST), Spintronics Research Center, Tsukuba, Japan
[3] Center for Nanoscale Science and Technology, National Institute of Standards and Technology, Gaithersburg, MD, USA
[4] Centre de Nanosciences et de Nanotechnologies, CNRS, Univ. Paris-Sud, Université Paris-Saclay, France

***Abstract*—** Fabricating powerful neuromorphic chips the size of a thumb requires miniaturizing their basic units: synapses and neurons. The challenge for neurons is to scale them down to submicrometer diameters while maintaining the properties that allow for reliable information processing: high signal to noise ratio, endurance, stability, reproducibility. In this work, we show that compact spin-torque nano-oscillators can naturally implement such neurons, and quantify their ability to realize an actual cognitive task. In particular, we show that they can naturally implement reservoir computing with high performance and detail the recipes for this capability.

## I. Introduction

Reservoir computing is a neural network-based theory, designed to process temporal inputs (Fig. 1) [1], [2]. A reservoir is composed of non-linear units, or neurons, that are connected recurrently through fixed connections (Fig. 1a). Input waveforms modify the activities of neurons inside the reservoir and the perturbed activities propagate through the recurrent network. The outputs of the reservoir are linear combinations of some or all neuron responses in the reservoir. The coefficients of these weighted sums are trained to obtain the desired outputs. A reservoir can classify waveforms due to the non-linearity of its neurons. It can also perform prediction tasks due to the recurrent connections which allow for fading memory of past inputs [3]. The limited number of connections to train make reservoir computing an excellent approach to evaluate novel technologies for neuromorphic computing.

In particular, we have recently shown experimentally that a single nanodevice, the spin-torque nano-oscillator, can implement reservoir computing through time multiplexing [4]. Spin-torque nano-oscillators (Fig. 2a) are magnetic tunnel junctions driven by dc current injection into a regime of sustained magnetization precession [5]. Magnetic oscillations are converted into voltage oscillations $V_{osc}$ through tunneling magneto-resistance (Fig. 2c). Despite the sub-micrometer diameter of such a spintronic artificial neuron, the experimental results for spoken digit recognition reach the state of the art for existing hardware and software (99.6 % recognition). In the present work, by analyzing a simple task, the classification of sine and square waveforms, we give a recipe for high classification performance with spin-torque nano-oscillators.

## II. Experimental Procedure

Our oscillators have FeB free layers with diameters of 375 nm, and a magnetic vortex as a ground state. A schematic of the experimental set-up is provided in Fig. 2b. We use the dc current to set the amplitude of voltage oscillations $\tilde{V}$ in the absence of inputs, and apply the input waveforms as a superimposed ac current. To compute, we use the fact that the amplitude of voltage oscillations across the junctions, $\tilde{V}$, is a non-linear function of the injected current ( Fig. 2d).

Using time-multiplexing, a reservoir can be emulated with a single oscillator, which plays the role of all neurons one after the other (Fig. 1b) [3]. This strategy requires that the state of this neuron at time $t+dt$ depends on its state at time $t$, just as a downstream neuron usually depends on the state of upstream neurons. This behavior can be achieved by pre-processing the input through multiplication with a binary fast-paced sequence that drives the oscillator into a transient state.

Figure 3 illustrates the procedure for reservoir computing. An input waveform, composed of randomly arranged sine and square waves with the same period, is shown in Fig. 3a [6]. The pre-processed input is displayed in Fig. 3b. It multiplies the input segment-wise with a binary sequence which has a total duration $\tau$ and is composed of $N_\theta$ points separated by a time interval $\theta$ ($N_\theta = \tau / \theta$). The number of points in the binary sequence, $N_\theta$, defines the size of the emulated network. For clarity, we have chosen in Fig. 3 to illustrate the working principle with a small neural network of only 12 neurons ($N_\theta$ = 12). The non-linear response of the amplitude of voltage oscillations $\tilde{V}$ to the pre-processed input is shown in Fig. 3c.

The response of the network is shown in Fig. 4a in a zoom-in of the response $\tilde{V}$ to a single input segment of duration $\tau$. Due to the oscillator non-linearity, each voltage amplitude value $\tilde{V}_i$ (i = 1…12) is a non-linear transform of the input value. In addition, $\tilde{V}_{i+1}$ depends on $\tilde{V}_i$. Indeed, magnetic oscillations have a relaxation time of about 300 ns, larger than the time interval $\theta$ = 100 ns between each $\tilde{V}_i$. The equivalent ring neural network is illustrated in Fig. 4c. The output of the neural network (Fig. 4b) is a sum of all $\tilde{V}_i$ : output = $\sum_{i=1}^{N_\theta} w_i \tilde{V}_i$ weighted by the strength of each connection, $w_i$ chosen to make the output match the target. The target for the output is a constant value for each waveform: one for squares, and zero for sines [6]. This output is reconstructed on a computer from the sampled experimental oscillator response $\tilde{V}$ by calculating the optimal weights through matrix inversion. Fig. 5 shows the reconstructed output obtained by experimentally emulating a 24-neuron network. The root mean square (rms) deviation between target and output is 11 %, which is small enough to distinguish between sines and squares without any error (perfect classification) for the chosen choice of parameters: dc current = 7.2 mA, magnetic field = 447 mT, input amplitude = 500 mV (equivalent to 6 mA).

## III. Results

The classification performances vary strongly depending on the experimental parameters. Indeed, the voltage amplitude $\tilde{V}$ (Fig. 6), the non-linearity $\partial^2\tilde{V}/\partial I^2$ (Fig. 7) and the voltage noise $\Delta V$ (Fig. 8), vary considerably with dc current and magnetic field. Since spin-torque oscillators have a small magnetic volume, thermal noise affects the magnetization dynamics. The resulting voltage amplitude noise is large for large non-linearity, which quantifies the sensitivity of the system to perturbations [7]. The correlation between voltage noise and non-linearity appears clearly in the comparison of Figs. 7 and 8. Neuron non-linearity is a key ingredient for classification as it allows the separation of input data [3]. On the other hand, noise in neuron response $\tilde{V}$ is detrimental for classification as it directly affects the output ( $\sum_{i=1}^{N_\theta} w_i\tilde{V}_i$).

Fig. 9 shows the classification performance as a function of dc current and field. We find good performance by choosing a bias point with intermediate non-linearity and therefore intermediate noise, and where the neuron output $\tilde{V}$ changes strongly in response to the ac input. Such bias points allow enough non-linearity to classify while keeping large enough signal to noise ratios to distinguish between outputs. As can be seen from Fig. 10, the larger the ac input variations (between 300 mV and 500 mV), the lower the rms deviations between output and target. Indeed, larger inputs lead to larger responses and improved signal to noise ratios.

Fig. 10 shows the evolution of the rms deviations between the output and the target as a function of the length of the time interval $\theta$ between $\tilde{V}$ samples. The evolution is completely different when the target is in exact phase with the input (Fig. 10a) and when the target is shifted by $\tau/2$ with respect to the input (Fig. 10b). Indeed, classification of sines and squares requires the network to have a short term memory of past inputs: some input points in the two different patterns are identical (+1 and -1 at their extrema). Therefore they can only be distinguished if the output of the network depends on the previous values of the input. When input and target are in phase (Fig. 10a), the only source of memory in the network comes from the relaxation time of the oscillator, of the order of 300 ns in our case [3]. This is why the classification performances degrade for $\theta$ > 300 ns in Fig. 10a. For sampling intervals that are too long, $\tilde{V}_{i+1}$ does not depend on $\tilde{V}_i$ anymore, and input sines and squares become difficult to separate. When $\theta$ is much smaller than the oscillator relaxation time, the oscillator cannot respond to the rapidly varying pre-processed input. Changes in $\tilde{V}$ become very small, the signal to noise ratio degrades and poor classification follows.

There is an optimum for the sampling interval $\theta$ (in our case $\theta_{opt}$ = 100 ns for ac input amplitudes of 500 mV). However, there is another way to endow the network with memory. Since the output is reconstructed offline after recording the whole response $\tilde{V}$ to inputs, it is possible to shift the target with respect to the input [8]. In that case, some of the samples $\tilde{V}_i$ used for reconstructing the output belong to the previous segment $\tau$. In other words, the current output is reconstructed partly from the present value of the input and partly from the last value of the input. The best results with this strategy, shown in Fig. 10b, are obtained when the number of samples is evenly distributed between the past and current input value. As in Fig. 10a, results are bad for small $\theta$ due to the low signal to noise ratio but they do not degrade for large $\theta$ values. The artificially introduced memory compensates for the loss of intrinsic memory.

## IV. Conclusion

Spin-torque nano-oscillators naturally provide the important features for implementing a reservoir computer. The recipe for high performance classification is the following: intermediate non-linearity and a high signal to noise ratio in the neural output. For tasks requiring short term memory, the intrinsic memory coming from magnetic relaxation times can be sufficient. When output reconstruction is done offline, an alternative strategy to endow the network with longer term memory is to shift the target with respect to the input. In the future, it will be interesting to introduce on-line long term memory through time-delayed feedback strategies.

## Acknowledgment

This work was supported by the European Research Council ERC under Grant bio*SPIN*spired 682955.

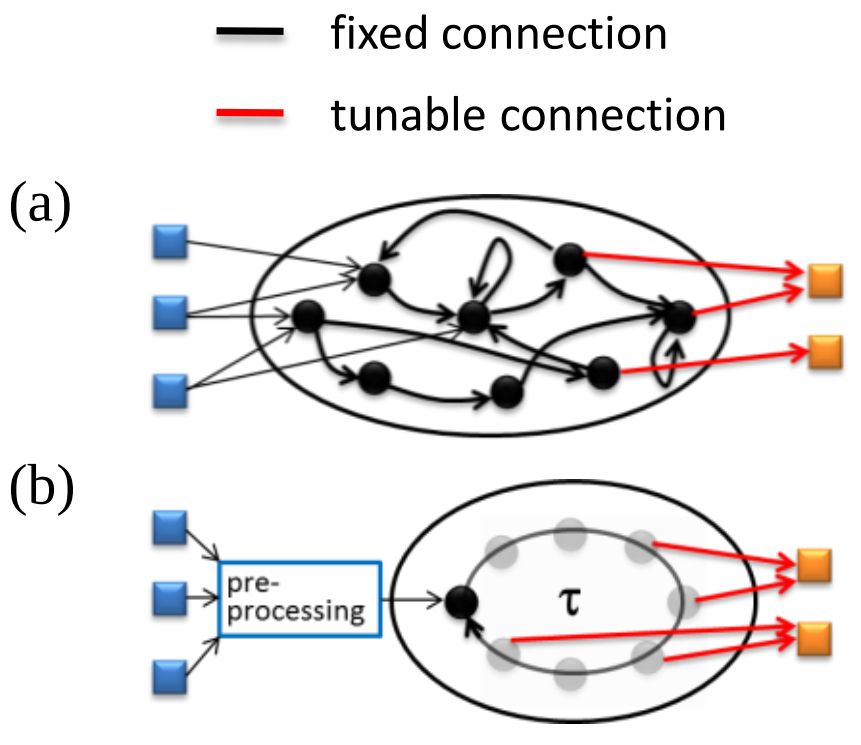

**Fig. 1.** **(a)** Illustration of reservoir computing concept. The network is composed by a large number of interconnected non-linear neurons. The internal connections are kept random and fixed and only external connections are trained. **(b)** Single neuron reservoir computing approach using time-multiplexing: the input is pre-processed in order to emulate neurons interconnected through time.

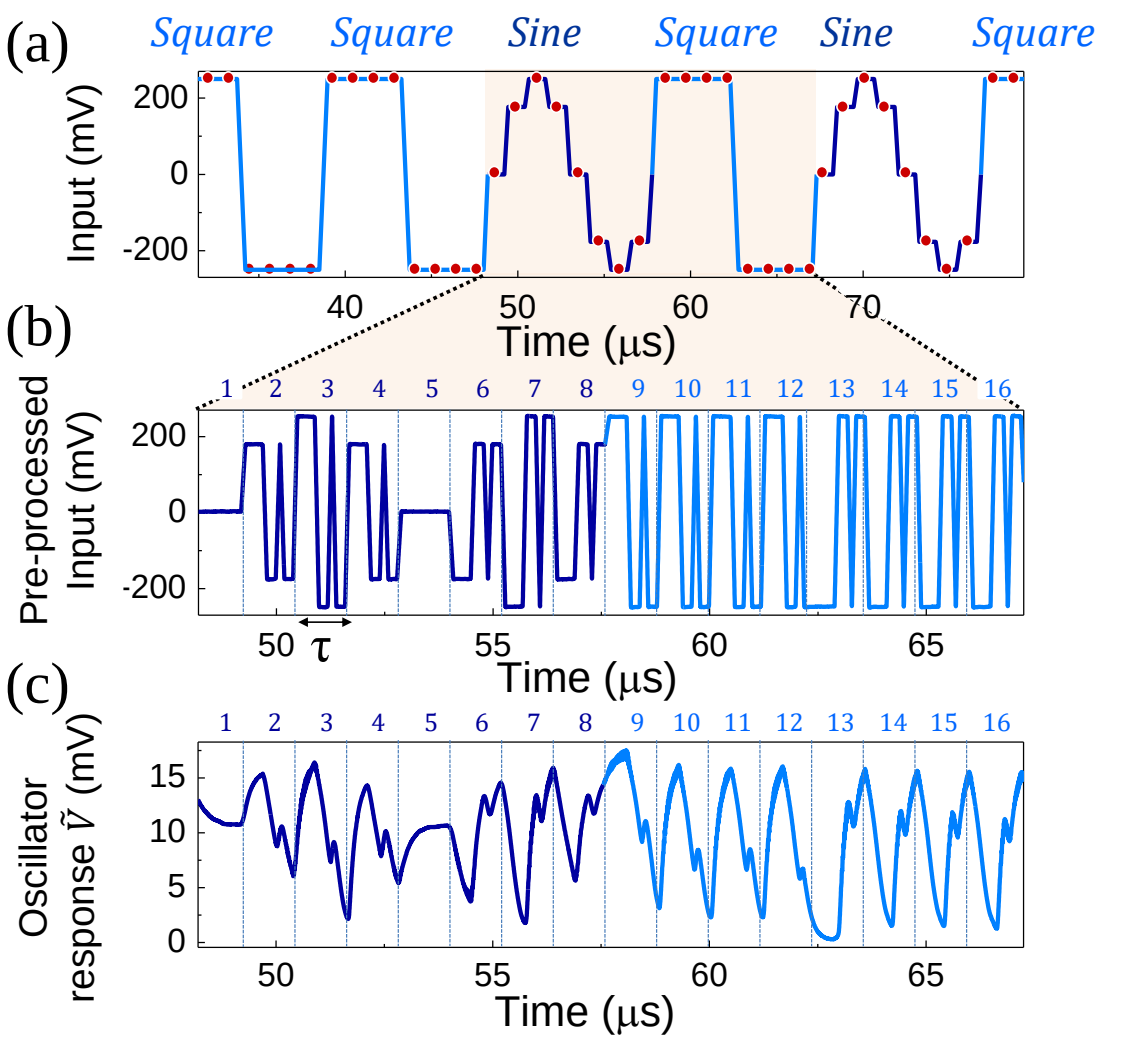

**Fig. 3.** **(a)** Input waveform. The task is to discriminate sines from squares at each red point. There are 8 discrete red points in each sine and square waveform. **(b)** Zoom-in on the preprocessed input waveform for a sine and a square. The corresponding fast binary input sequences are numbered from 1 to 16 (8 for the sine, 8 for the square. **(c)** Envelope $\tilde{V}(t)$ of the experimental oscillator emitted voltage amplitude ($\mu_0 H$ = 466 mT, $I_{DC}$ = 7 mA). The trajectories created in response to the input waveform are numbered from 1 to 16 (8 for the sine, 8 for the square).

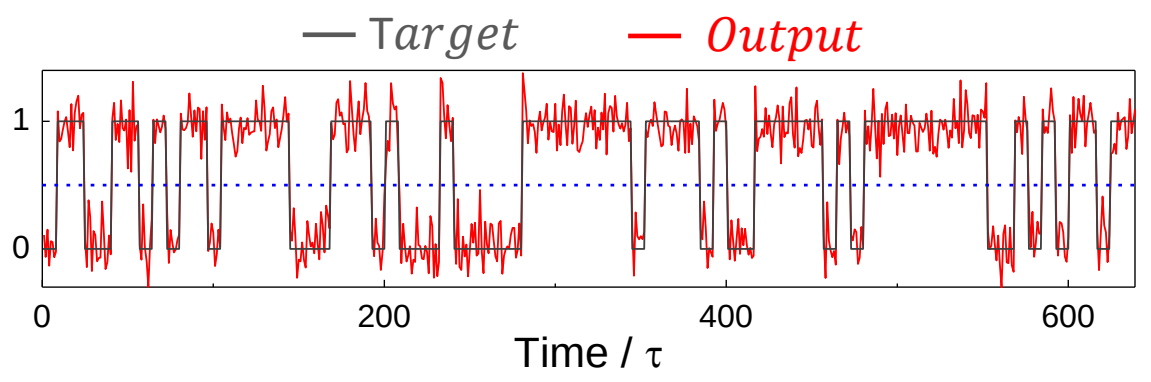

**Fig. 5.** Reconstructed output (red) and target (grey) in response to an input waveform with 80 randomly arranged sines and squares. The magnetic field is $\mu_0 H$ = 447 mT, and the applied current 7.2 mA. The results are based on 24 neurons separated by θ = 100 ns.

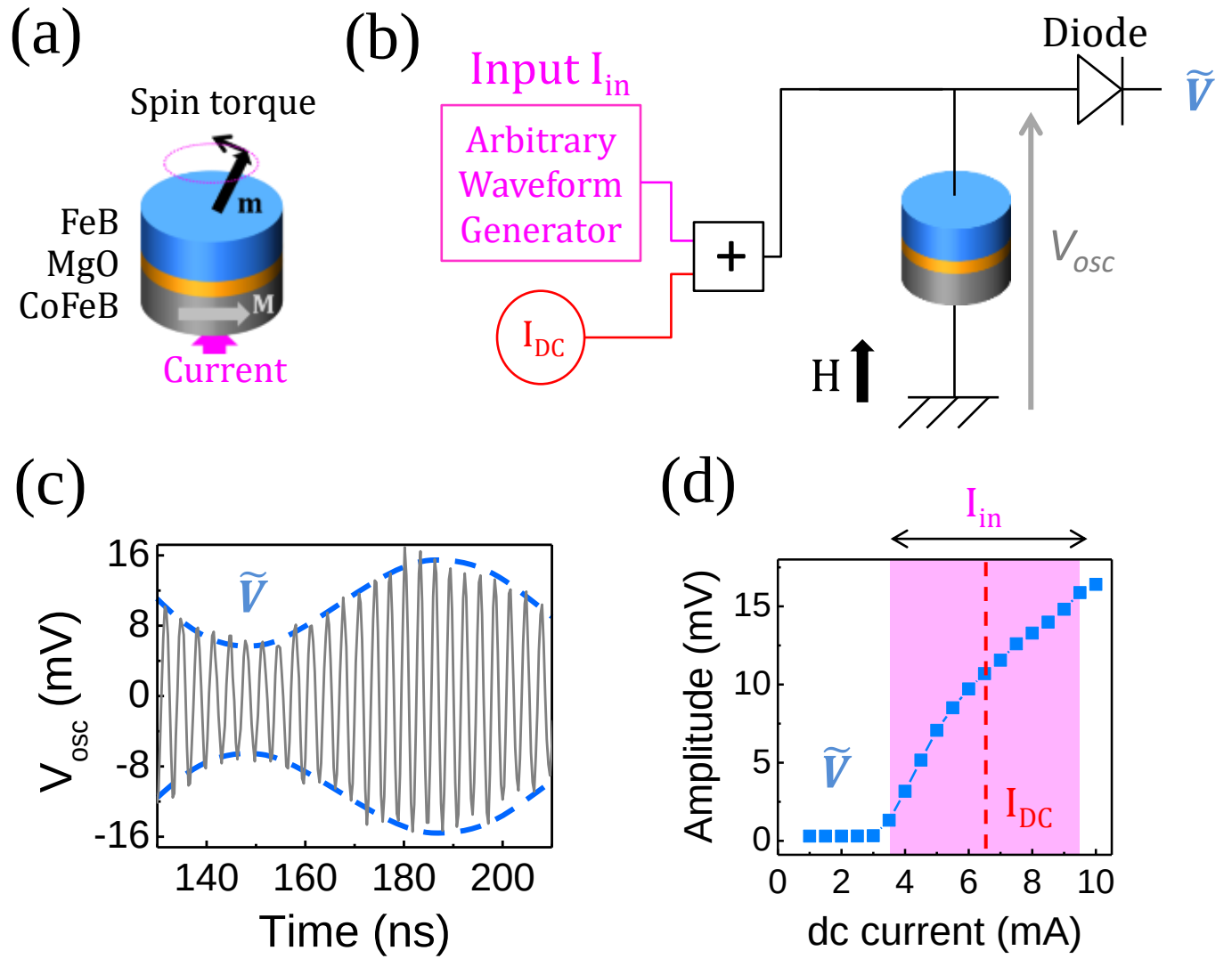

**Fig. 2.** **(a)** Spin-torque nano-oscillators used for neuromorphic computing: magnetic tunnel junction (CoFe/MgO/FeB) driven by spin transfer torque. **(b)** Schematic of the experimental set-up. A dc current $I_{DC}$ as well as a fast-varying waveform encoding the input $I_{in}$ are injected in the spin-torque nano-oscillator. **(c)** The microwave voltage $V_{osc}$ emitted by the oscillator is measured with an oscilloscope. For computing, the amplitude $\tilde{V}$ of the oscillator is used, and measured directly with a microwave diode. **(d)** Voltage amplitude $\tilde{V}$ as a function of current at $\mu_0 H$ = 430 mT. The typical resulting excursion of current amplitude is highlighted in magenta when an input signal with maximum amplitude ±3 mA (corresponding to ±250 mV), is injected.

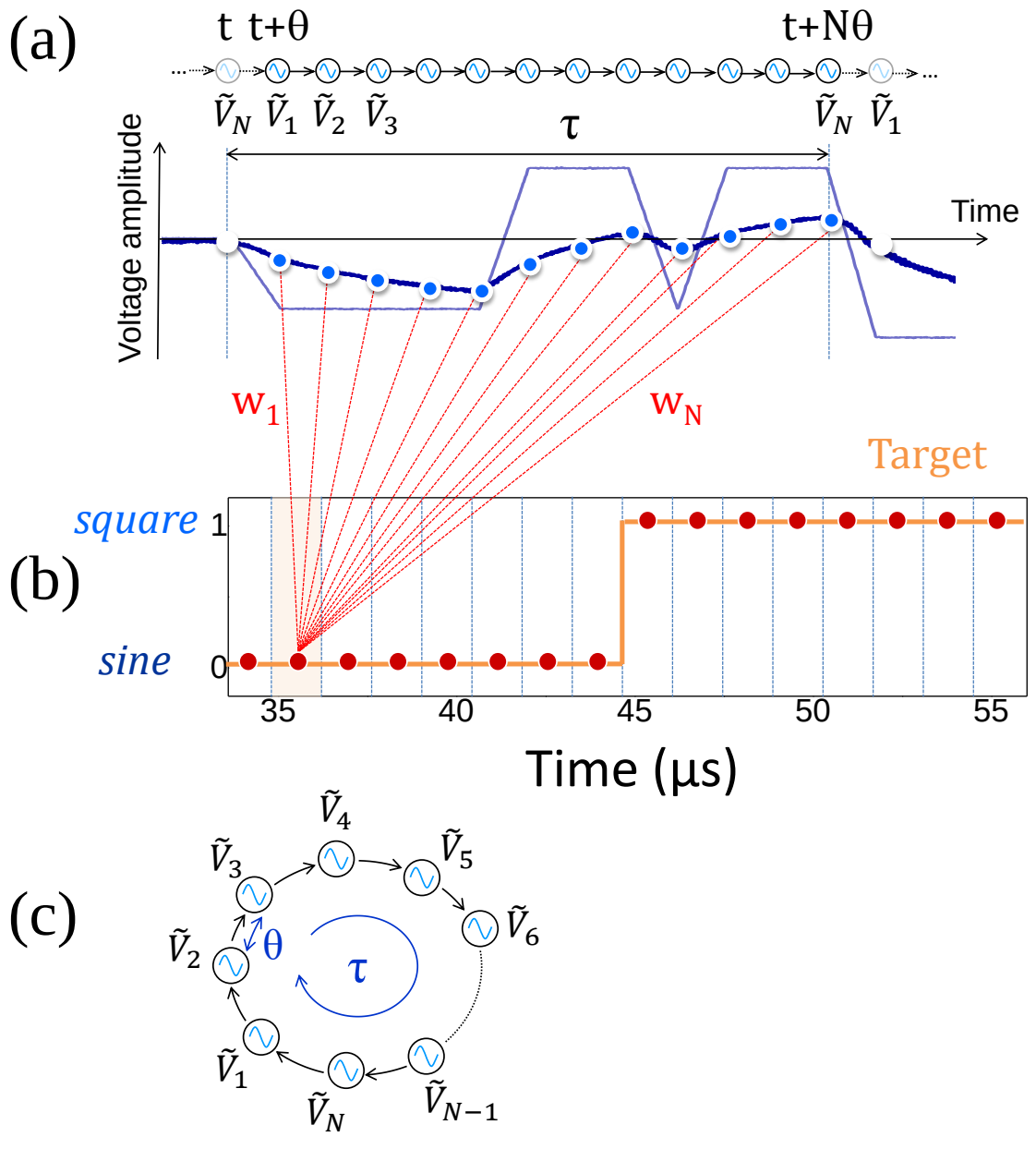

**Fig. 4.** **(a)** Oscillator voltage amplitude $\tilde{V}$ changes corresponding to a single time segment $\tau$. Here, 12 neurons (12 samples $\tilde{V}_i$ separated by the time step θ) are used to construct the output. **(b)** The transient states of the oscillator give rise to a chain reaction emulating the neural network with a ring structure **(c)** Target for the output reconstructed from the voltages $\tilde{V}_i$ in each time segment $\tau$: output = $\sum_{i=1}^{N} w_i \tilde{V}_i$.

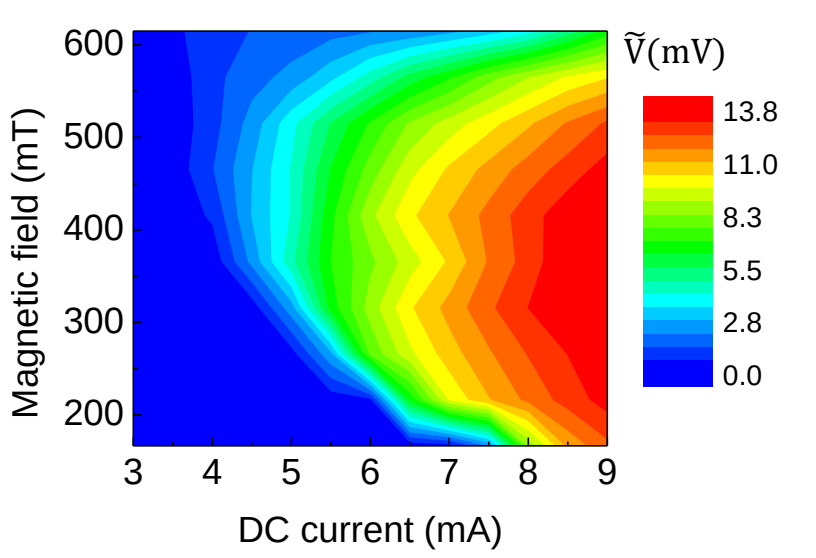

**Fig. 6**. Amplitude Voltage $\tilde{V}$ of the oscillator in the steady state: map in the $I_{DC}$ - $\mu_0 H$ plane.

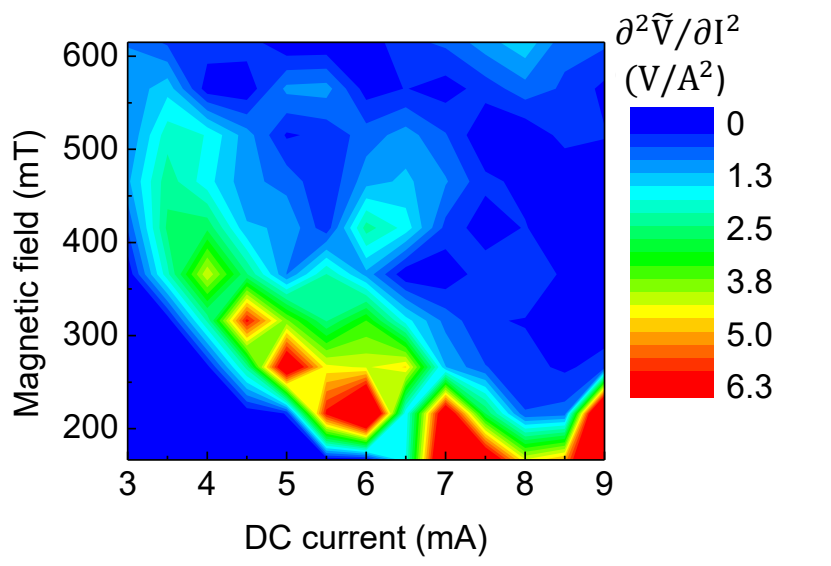

**Fig. 7**. The non-linearity $\partial^2\tilde{V}/\partial I^2$ of the oscillator: map in the $I_{DC}$ - $\mu_0 H$ plane.

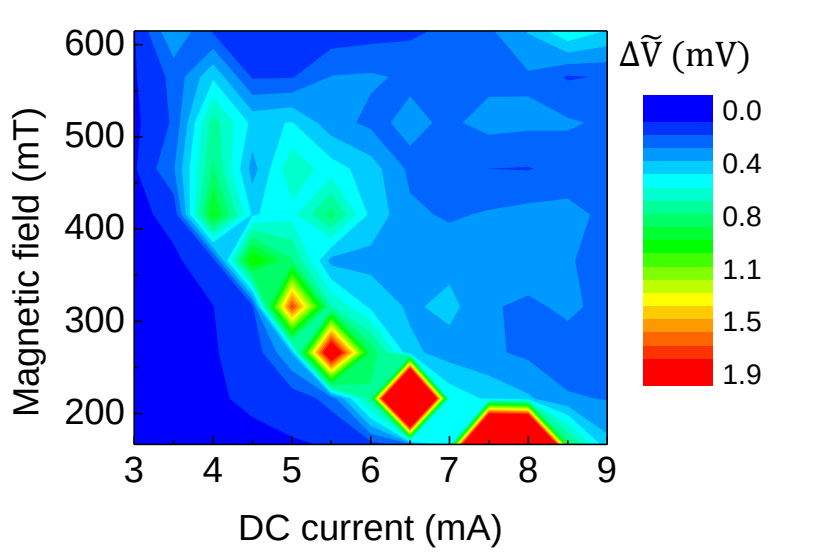

**Fig. 8**. Amplitude noise $\Delta\tilde{V}$ of the oscillator in the steady state: map in the $I_{DC}$ - $\mu_0 H$ plane.

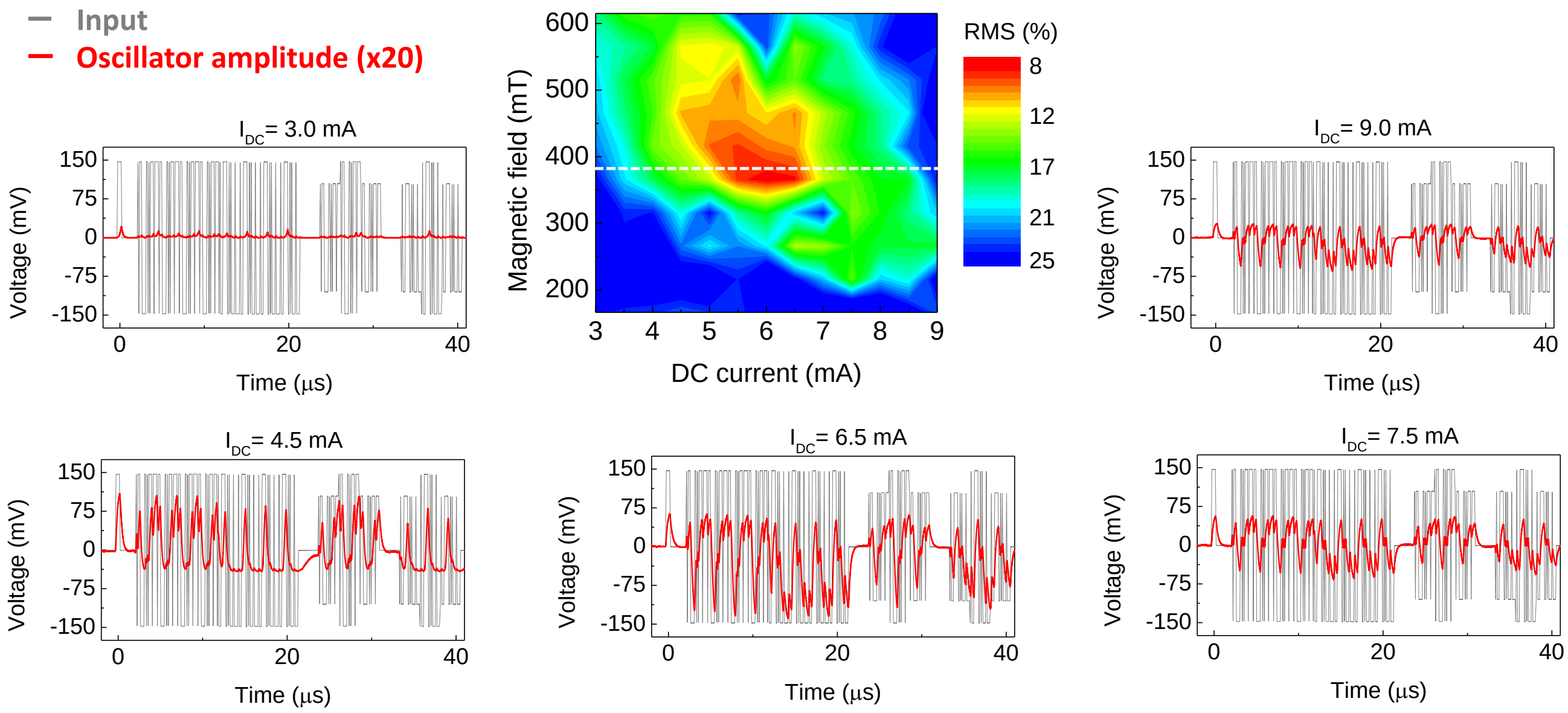

Fig. 9. Root mean square of output-to-target deviations: map as a function of dc current $I_{DC}$ and magnetic field $\mu_0$H. The oscillator voltage amplitude curves (in red) in response to the input waveform (in gray) are plotted for selected dc currents (3, 4.5, 6.5, 7.5 and 9) mA and magnetic field $\mu_0 H$ = 380 mT. Here $V_{in}$= 300 mV and $\theta$= 100 ns is used. RMS map corresponds to the target shifted by $\tau/2$ with respect to the input.

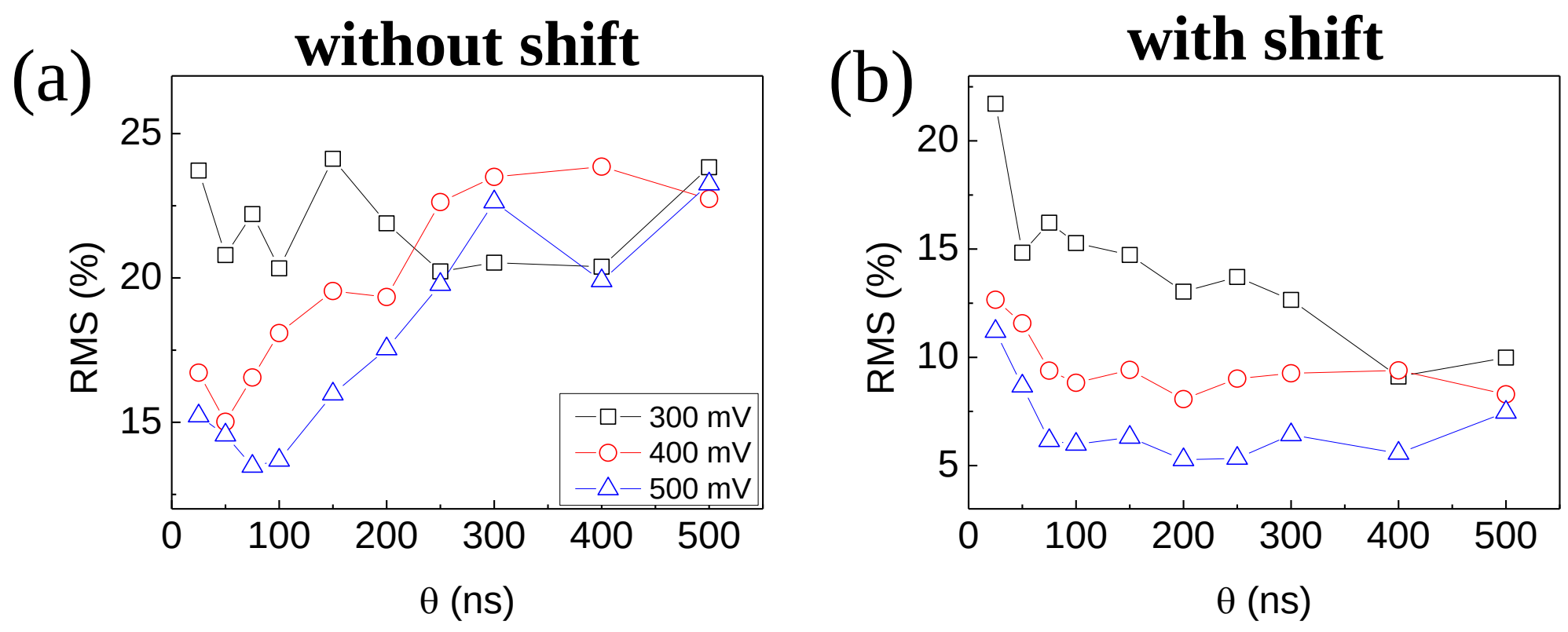

**Fig. 10.** Root mean square of output-to-target deviations as a function of the time step θ (separation between transient states of the oscillator $\tilde{V}_i$) for different amplitudes of the input signal (300, 400 and 500) mV: **(a)** the target is in exact phase with the input and **(b)** the target is shifted by $\tau/2$ with respect to the input.